\documentstyle[epsf,floats,preprint,aps,epsfig]{revtex}

\input epsf
\tighten
\def\bgamma{\mbox{\boldmath $\gamma$}}
\def\br{\mbox{\boldmath $r$}}

\def\mph{m_{\phi}}
\def\mpi{m_{\phi}^{-1}}

\def\mwi{m_W^{-1}}

\def\high{\vphantom{\Biggl(}\displaystyle}

\begin{document}

\newcommand{\be}{\begin{equation}}
\newcommand{\ee}{\end{equation}}
\newcommand{\bea}{\begin{eqnarray}}
\newcommand{\eea}{\end{eqnarray}}

\rightline{CU-TP-1077}
\rightline{hep-th/0211050}
\vskip 1cm

\begin{center}
\ \\
\large{{\bf Massless monopole clouds and electric-magnetic duality}} 
\ \\
\ \\
\ \\
\normalsize{ Xingang Chen\footnote{Current address: Institute for
Fundamental Theory, Department of Physics, University of Florida,
Gainesville, FL 32611} } 
\ \\
\ \\
\small{\em Physics Department, Columbia University \\ New York,
New York 10027} 

\end{center}

\begin{abstract}
We discuss the Montonen-Olive electric-magnetic duality
for the BPS massless monopole clouds in ${\cal N}=4$ supersymmetric
Yang-Mills theory with non-Abelian 
unbroken gauge symmetries. We argue that these low energy non-Abelian
clouds can be identified as the duals of
the infrared bremsstrahlung radiation of the
non-Abelian massless particles. After we break the ${\cal N}=4$
supersymmetry to ${\cal N}=1$ by adding a superpotential, or to ${\cal
N}=0$ by further adding soft breaking terms, these non-Abelian clouds
will generally condense and screen the non-Abelian charges of the
massive monopole probes. The effective mass of these dual non-Abelian
states is likely
to persist as we lower the energy to the QCD scale, if all the
non-Abelian Higgs particles are massive. This can be regarded as a
manifestation of the non-Abelian dual Meissner effect above the QCD
scale, and we expect it to continuously connect with the confinement
as we lower the supersymmetry breaking scale to the QCD scale.
\end{abstract}

\setcounter{page}{0}
\thispagestyle{empty}
\maketitle

\eject

\vfill

\baselineskip=18pt

\section{Introduction}
The ${\cal N}=4$ supersymmetric Yang-Mills theory is conjectured to
have a remarkable electric-magnetic duality
\cite{Montonen:1977sn,Osborn:tq,Sen:1994yi,Girardello:qt}.
A special form of this 
conjecture suggests that the electric theory
is dual to the magnetic theory with a dual group and an inverse
coupling constant. 

This conjecture originated in the study of an SU(2) theory
spontaneously broken to U(1)\cite{Montonen:1977sn,Osborn:tq}, where
there is only one type of fundamental 
(anti-)monopole. The supersymmetric multiplet based on this monopole
is dual to the massive gauge supermultiplet.

\begin{figure}[t]
\begin{center}
\epsfig{file=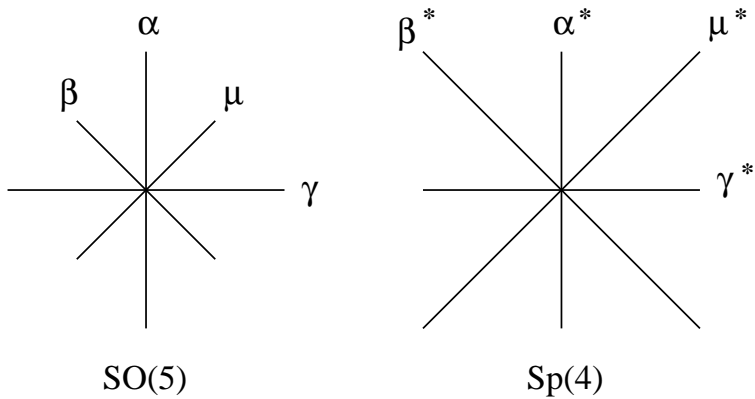, width=10cm}
\end{center} 
\medskip
\caption{The root diagrams of SO(5) and Sp(4).}
\label{roots}
\end{figure}

If the rank $r$ of the gauge group is higher than one, when the
non-Abelian gauge symmetry is maximally broken to U(1)$^r$, the
monopole configurations can be treated as superpositions of 
fundamental monopoles associated with simple roots
\cite{Weinberg:1979zt}; while for elementary particles,
each root of the dual group corresponds to a massive gauge
supermultiplet. Studies of supersymmetric sigma models
on the monopole 
moduli spaces \cite{Gauntlett:1996cw} 
show that these 
supersymmetric fundamental 
monopoles indeed form threshold bound states as predicted by the
duality. To illustrate this, we
use monopoles in SO(5) $\rightarrow$ U(1)$^2$ 
theory. The root diagrams of SO(5) and its dual group Sp(4) are
shown in  
Fig.\ref{roots}. 
A unique normalizable anti-self-dual harmonic two-form is found
\cite{Gauntlett:1996cw} on the moduli space of the {\boldmath
$\beta$} and 
{\boldmath $\gamma$} monopoles (we have chosen and labeled the simple
roots as 
{\boldmath $\beta$} and {\boldmath $\gamma$}). 
This corresponds to a threshold bound state. 
The supersymmetric multiplet associated with this bound
state in SO(5) is dual to the 
{\boldmath $\alpha^*$} gauge 
multiplet in Sp(4) as predicted by the duality.

The situation is more subtle when
a non-Abelian subgroup of the gauge symmetry remains unbroken. In
such cases, 
BPS configurations 
of massless monopole clouds (or non-Abelian clouds) have been found
\cite{Weinberg:jh,Lee:1996vz}. These configurations describe massive
monopoles surrounded by clouds of the non-Abelian fields, which form
an overall magnetic color singlet. In addition
to the usual moduli of positions and U(1) phases of the massive
monopoles, there are also ones describing the unbroken non-Abelian
gauge group, as well as the sizes or shapes of the clouds. There have
been many extended studies on both the BPS configurations
\cite{Dancer:kn,Irwin:1997ew,Lee:1997ny,Weinberg:1998hn,Lu:1998br,Houghton:2002bz,Chen:2002vb}
and the low
energy classical dynamics \cite{Dancer:hf,Irwin:2000rc,Chen:2001qt} of
such clouds. However, it remains unclear how such 
configurations and their properties should be properly placed in the
context of 
the electric-magnetic duality of the ${\cal N}=4$ theory and how it
may be related to the properties of the QCD confinement. It is the
purpose of this paper to make some initial steps toward this
direction. First we observe that these low energy non-Abelian clouds
should be identified as the dual bremsstrahlung radiation of the
non-Abelian massless particles. Then, after breaking the
supersymmetry,
we argue that the non-BPS properties of these cloud are the
manifestation of the non-Abelian dual Meissner effect at weak electric
coupling 
above the QCD scale. We expect it to continuously go to the
non-Abelian dual Meissner
effect in QCD confinement when we lower the supersymmetry breaking
scale to the QCD scale.

\section{Massless monopole clouds and bremsstrahlung radiation}
\label{SecDuality}
We use the same SO(5) example.
The gauge symmetry is now partially broken to
SU(2) $\times$ U(1) 
by a Higgs expectation value {\boldmath $h$} orthogonal to the root
{\boldmath 
$\gamma$} or {\boldmath $\gamma^*$}
\cite{Weinberg:ev}. Correspondingly, the $\bgamma$ monopoles 
or $\bgamma^*$ elementary particles become massless. 
A spherically symmetric 
BPS magnetic monopole solution is found in \cite{Weinberg:jh}. It
describes a  
massive monopole, embedded in the SU(2) subgroup defined by the root
{\boldmath $\beta$}, surrounded
by a 
non-Abelian cloud. There is a modulus $a$ characterizing the
size of 
the cloud.  We will be interested only in the
non-Abelian
fields which do not exponentially decay outside the massive monopole
core $m_{W}^{-1}$: 
\be
A^a_{i( {\bf \gamma} )} = \epsilon_{aim} {\hat r}_m G(r)\, , \qquad
\phi^a_{ {\bf \gamma} }={\hat r}_a G(r)~,
\label{naf}
\ee
where the subscripts {\boldmath $\gamma$} mean that the fields
correspond to the triplet SU(2) generators 
$t_a$({\boldmath $\gamma$}) $(a=1,2,3)$ associated with the root
{\boldmath $\gamma$},
and
\be
G(r)=\frac{1}{e r(1+r/a)} ~.
\label{Gform}
\ee
If the cloud size $a$ is infinite, we only have the 
massive monopole,
carrying both Abelian and non-Abelian charges. If $a$ is finite,
the cloud shields the non-Abelian charge of 
the massive one, so that the non-Abelian fields fall as $a/r^2$
outside of the radius $a$, as we can see from (\ref{naf}) and
(\ref{Gform}).

The metric for this massless monopole cloud can be obtained 
\cite{Lee:1996vz} by taking
the zero reduced mass limit of the maximally broken case: 
\be
ds^2 = \frac{g^2}{8\pi} \left( \frac{da^2}{a} + a \sigma_1^2 + a
\sigma_2^2 + a \sigma_3^2 \right) ~,
\label{cloudmetric}
\ee
where $g=4\pi/e$ is the magnetic coupling and $\sigma_i$ $(i=1,2,3)$
are the one-forms describing the unbroken SU(2). For this metric,
the harmonic (anti-)self-dual form 
is not normalizable. So the massless monopole cloud is not
bound. It has 
been a puzzle \cite{Lee:1996vz} why this configuration, which is dual
to the {\boldmath $\alpha^*$} gauge multiplet in the Sp(4) theory,
does not have a 
normalizable threshold bound state as in the maximally broken
case. 

To answer that, we first look at the elementary particles in the
weakly coupled electric theory of Sp(4).
Because the beta function vanishes in 
the ${\cal N}=4$ supersymmetric gauge theory \cite{Sohnius:sn}, the
massless particles of the non-Abelian gauge multiplet {\boldmath
$\gamma^*$} in this weakly coupled theory are not 
confined. Therefore, whenever a massive particle is coupled to these
massless 
ones, it emits non-Abelian infrared bremsstrahlung radiation. For
example, the massive {\boldmath $\alpha^*$} Higgs can become a massive
{\boldmath $\beta^*$} Higgs by emitting an infrared
gauge or 
Higgs boson associated with the root {\boldmath
$\gamma^*$}. Generalizing this, the 
massive gauge multiplets {\boldmath $\alpha^*$} 
and {\boldmath $\beta^*$} become indistinguishable
through the emission and 
absorption of the massless gauge supermultiplet associated
with {\boldmath $\gamma^*$}.\footnote{It is
interesting to compare this SO(5) example to a single massive
fundamental monopole in  SU(3) $\rightarrow$
SU(2) $\times$ U(1) theory, where the massless monopole cloud is
absent. On the dual side, for a single massive 
elementary particle in this SU(3) theory, the non-Abelian charge is
unchanged (or gauge equivalent) after infrared radiation.}

These two descriptions for the monopoles and elementary particles
are very different. The former describes a solitonic static field
configuration, while the latter describes massless elementary
particles that propagate in the speed of light. To see how
they can be dual to each other, we need to analyze
the low energy supersymmetric quantum mechanics
of the massless monopole clouds on the moduli space.

To see what happens, we need to
find the spherically symmetric eigenstates of
the Laplacian $\triangle = d d^\dagger + d^\dagger d$
corresponding to the metric (\ref{cloudmetric})
\cite{Witten:df}. These non-normalizable scattering states can be
described by sixteen harmonic
differential forms which are the duals of the gauge supermultiplet
$\bgamma^*$. Up to constant factors, these are given by
\bea
{\rm 0-form:} ~~~~ &{\high \frac{1}{\sqrt{a}} J_1 \left( g\sqrt{Ea/2\pi}
\right) }~, \label{CloudZeroForm} \\
{\rm 1-forms:} ~~~~ &{\high \frac{1}{a} J_2 \left( g\sqrt{Ea/2\pi}
\right) da }~, ~~~~
J_2 \left( g\sqrt{Ea/2\pi} \right) \sigma_i ~, ~~ (i=1,2,3) ~, \\
{\rm 2-forms:} ~~~~ &{\high \frac{1}{\sqrt{a}} J_1 \left
( g\sqrt{Ea/2\pi} \right) \left(da \wedge \sigma_1 + a \sigma_2 \wedge
\sigma_3 \right) }
~~, {\rm ~and ~ cyclic,} \\
& {\high \frac{1}{\sqrt{a}} J_3 \left( g\sqrt{Ea/2\pi} \right)
\left(da \wedge \sigma_1 - a \sigma_2 \wedge \sigma_3 \right) } 
~~, {\rm ~and ~ cyclic,} \label{CloudTwoForm}
\eea
where $E$ is
the arbitrarily small energy of 
the massless monopole cloud and $J$'s are Bessel functions.
The $\sigma_i$'s and $da$ correspond to fermionic excitations.
The 3-forms and 4-form are the Hodge duals of the 1-forms and 0-form,
respectively. 
The $a$-dependence of these wave functions are all
similar. For example, the 0-form
wave function goes to a constant for $a \ll \frac{2\pi}{g^2 E}$ and
falls as $a^{-3/4} \cos(g\sqrt{Ea/2\pi}-3\pi/4)$ for $a \gg
\frac{2\pi}{g^2 E}$.

\begin{figure}[t]
\begin{center}
\epsfig{file=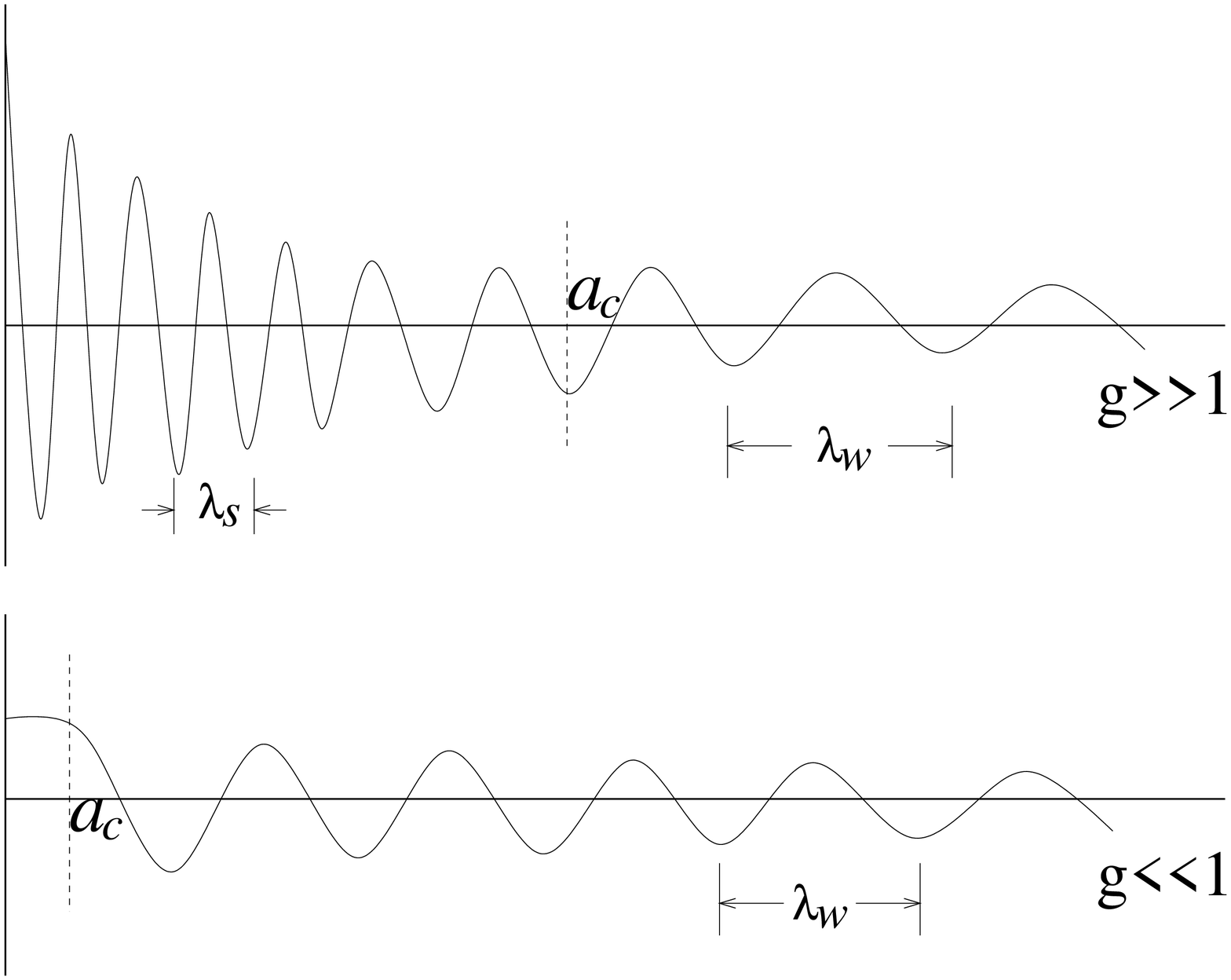, width=9cm}
\end{center} 
\medskip
\caption[The cloud wave functions]{The cloud wave functions for the
two different cases, $g\gg
1$ and $g \ll 1$. The three different length scales $a_c\approx
g^2/E$, $\lambda_w \approx 1/E$ and $\lambda_s \approx 1/(g^2 E)$ are
indicated in these two different cases. The part within $a_c$ is the
solitonic phase, whose wave function is given by
Eqs.~(\ref{CloudZeroForm}) - (\ref{CloudTwoForm}). Outside $a_c$ is
the spherical wave. For $g \gg 1$, $a_c$ is the
biggest length scale, while for $g \ll 1$, $a_c$ is the smallest. }
\label{cloudwave}
\end{figure}

However the moduli space approximation for the low energy solitons
usually requires small velocities. For the case of the massless
monopole cloud, this requires \cite{Chen:2001qt} $\dot a <1$. From
the metric (\ref{cloudmetric}), this imposes
the restriction $a<a_c=\frac{g^2}{8 \pi E}$. Beyond this region the
moduli space approximation fails and the cloud propagates as a
wavefront at the speed of light \cite{Chen:2001qt}. So the
wave function should be replaced by the spherical wave $\sim
e^{iEa}/a$ as $a>a_c$, where $a$ becomes the
position of the wavefront. As we turn to the
weak magnetic coupling (small g) limit, the duality conjecture
suggests that the monopoles and the elementary gauge particles
exchange roles. Indeed, as we can see,
the extent of the solitonic wave
function $a_c$ is much smaller then the wavelength ($\sim 1/E$) of
the wavefront and, in addition, inside $a_c$ the wave function is
nearly
a constant. Thus the solitonic phase is negligible. (See
Fig.~\ref{cloudwave}.) The massless
monopole always 
appears as infrared radiation and the elementary local field
description takes over.

The above discussion is in accordance 
with the classical dynamics discussed in\cite{Chen:2001qt},
i.e.~the prediction of the moduli space approximation from
(\ref{cloudmetric}) that $a \sim t^2E/g^2$
is good only for a time period of order $g^2/E$ during which the cloud
speed $\dot a
<1$. According to the uncertainty principle, for $g<1$, it is quantum
mechanically unobservable.

So instead of being a problem, the absence of the massive and massless
monopole bound states is in fact consistent with the duality. As we
turn the coupling $g$ from strong to weak, the unbound cloud becomes
the infrared non-Abelian radiation of elementary particles and the
solitonic phase of the cloud disappears quantum
mechanically.\footnote
{\label{ftHiggs}Since we are considering the case where all six Higgs
vevs in the
${\cal N}=4$ theory are proportional in the gauge space, a
global SO(6) rotation
can make all but one of them zero.  
An important difference when we go from big $g$ to small $g$ is that,
while the Higgs profile of the strongly coupled non-Abelian massless
monopole clouds take non-zero 
values only in one Higgs direction, as we go to the weak
coupling limit, we can see from the interaction terms in the field
theory that these massless radiations can oscillate in all the Higgs
directions. This fact will also be useful later.}

In the above discussion, we always studied the weakly coupled theory
in the 
elementary particle sector where particles are local excitations of 
fields, and the strongly coupled theory in the solitonic sector where
solitons are non-local objects from the point of view of the elementary
fields. Both descriptions can happen either in the electric or
magnetic theory, depending on the couplings. We will continue to use
this method throughout the paper.

\section{Dynamics of non-BPS non-Abelian monopole clouds}
Non-Abelian ${\cal N}=1$ or ${\cal
N}=0$ supersymmetric gauge theories have the important property of
confinement. Significant insights have been made  
by Seiberg and Witten in \cite{Seiberg:1994rs}. From the exact ${\cal
N}=2$ low 
energy theory, they explicitly show that a superpotential breaking the
supersymmetry to ${\cal N}=1$ causes the massless magnetic monopole
field to 
condense. This confinement is described in a weakly coupled
magnetic theory through the
dual Meissner effect
\cite{Mandelstam:1974pi}. Related
issues starting 
from ${\cal N}=4$ have also been studied (see
e.g.\cite{Strassler:1997ny} and references therein).

It is natural to ask what roles the non-Abelian
clouds we have studied may play in
this QCD confinement. To see this, we will focus on the energy region
above the QCD scale $\Lambda_{QCD}$. Specifically, we start with a
${\cal N}=4$ theory with a weak electric coupling at high energy. In
this theory we have argued that,
in the presence of certain massive monopoles, we
can identify the low energy
magnetic non-Abelian clouds as 
the dual infrared non-Abelian particles by exploring the
duality conjecture. When we break the supersymmetry at low energy, we
break the original electric-magnetic symmetry. But the
dual states we 
identified should still exist and we will be interested in how they
evolve as the supersymmetry is broken. As mentioned, we will focus
mostly on the 
energy region above $\Lambda_{QCD}$, where the strongly
coupled magnetic theory is described by non-BPS monopoles. Then we
will discuss some implications for the low energy theory below
$\Lambda_{QCD}$. 
 
We explicitly break the
supersymmetry to ${\cal N}=1$ at low energy by adding a
superpotential for the ${\cal N}=1$ chiral multiplets.
We expand the Higgs around those vacua where part of the non-Abelian
symmetry is 
unbroken and use $\phi$
to represent the non-Abelian components of the
deviations. Among all the terms in the expansion, 
we will study the quadratic terms
\begin{equation}
\frac{1}{2} m_{\phi}^2 {\rm tr}(\phi^2)
\label{potential}
\end{equation}
as examples. This gives an ${\cal N}=4$ supersymmetry scale $\mph$. As
mentioned, the fact that $m_{\phi} > \Lambda_{\rm QCD}$ is guaranteed
as long as the electric coupling is weak at the supersymmetry breaking
scale $m_{\phi}$. We
will be 
interested in the limit where the non-Abelian 
Higgs masses $m_{\phi}$ are much smaller than the massive
gauge bosons 
$m_W$. We also want the U(1) Higgs masses
to be much smaller than the non-Abelian Higgs. By doing
this, we effectively make the U(1) parts remain BPS so we can
concentrate on 
the non-BPS properties of the non-Abelian parts only. This is why we
have neglected the U(1) mass terms in (\ref{potential}).\footnote{
In certain models, the above mass relations can be achieved by
adjusting the parameters 
in the potential. 
For instance, consider an SU(2$N$) theory with superpotential
\be
W(\Phi) = m~{\rm tr}(\Phi^2) + \lambda~{\rm tr}(\Phi^3)+\eta X
\left( {\rm tr}(\Phi^2) - \mu^2 \right) ~,
\label{su2nsp}
\ee
where $m$, $\mu$, $\lambda$ and $\eta$ are real, and we
have introduced a color singlet $X$ to have more adjustable
parameters. At $X = -m/\eta$,
this theory has a ${\cal N}=1$ supersymmetric vacuum where the gauge
symmetry is broken to SU($N$) $\times$ U(1) $\times$ SU($N$). The mass
relations can be satisfied by choosing
$e \gg \lambda/\sqrt{N} \gg \eta$. Also consistent with this
requirement, the dimensionless couplings in this superpotential have
to be very small comparing to the electric coupling $e$ at the
supersymmetry breaking scale,
so that above this scale the ${\cal N}=4$ supersymmetry is
restored. [Because the dimensionless parameters $\lambda$ and
$\eta$ grow
when we increase the energy, at much higher energy we again return to
${\cal N}=1$. But this does not affect our argument as long as there
is a region where ${\cal N}=4$ supersymmetry is approximately held.]
However to illustrate the properties of the non-Abelian clouds,
we will use simpler groups such as the previously mentioned
SO(5). We will not try to construct the specific potential for each
case, because the simple 
qualitative features which will be summarized after those examples are
true for cases where these mass relations are satisfied.}

To study the
non-BPS monopoles, it is enough to add a superpotential in the
direction of the non-zero Higgs. But for later purposes to connect
with confinement, we will also add superpotentials for the other
two chiral multiplets. This can
be simply given by the mass terms with zero Higgs vev. It has no
effect 
on the monopole properties we will discuss. 

We first study the example in SO(5). The BPS fields are given in
(\ref{naf}).
When the non-BPS potential (\ref{potential}) is added, the non-Abelian
Higgs field is
exponentially cut off at a distance scale $m_{\phi}^{-1}$. Outside
of the region $m_{\phi}^{-1}$ where the
Brandt-Neri-Coleman (BNC) instability \cite{Brandt:kk,Coleman:1982cx}
applies, the gauge field 
decays to a magnetic-color 
neutral configuration, which corresponds to having a non-Abelian cloud
inside $\mpi$. Since
$m_{\phi}^{-1} \gg \mwi$, the BPS solution (\ref{naf}) is still a good
approximation between $m_{\phi}^{-1}$ and $\mwi$. However, the cloud
size is no longer  
a modulus. It is easy to see that, under the potential
(\ref{potential}), it is classically energetically favored for the
cloud to shrink.
We can use the BPS solution to estimate this $a$-dependent
potential. It is\footnote{Note we have a non-standard kinetic term for
$a$ from Eq.~(\ref{cloudmetric}).}
\be
\frac{g^2}{8\pi} m_{\phi}^2 a ~.
\label{apotential}
\ee
This should be a good approximation
as the non-BPS potential is weak. The
correction is given by factors of $m_{\phi} a$. The 
potential change within 
the core, $r < \mwi$, is negligible.

Using the metric (\ref{cloudmetric}) and this linear potential, we
can study the quantum  
mechanics of this bounded non-Abelian cloud. This is
non-supersymmetric, as the monopole breaks the ${\cal N}=1$
supersymmetry. For the purpose of this paper,
we simply note that the ground state of the cloud has a 
mass gap of order $\mph$ and is concentrated in the region
$\langle a \rangle \sim g^{-2}\mpi \ll \mpi$, since the factor $g^2$
can be absorbed
in the $a$ in (\ref{cloudmetric}) and (\ref{apotential}).
Any multi-monopole configuration can be thought of 
as being a
collection of these color singlets. Since we neglected the U(1)
Higgs mass, there are no net long-range forces between the monopoles
when they 
are separated further than $\mpi$. 
Before discussing the physical interpretation of this result, we
consider a case where the cloud 
encloses two 
massive monopoles. 

We use the
minimal symmetry breaking model of SU(3) \cite{Dancer:kn}. When the
two massive 
monopoles are far apart, so that the non-Abelian Higgs has decayed
exponentially, 
the relative orientation of their non-Abelian 
gauge charges is
self-adjusted to minimize the energy \cite{Coleman:1982cx}. The
charges are then given by  
\be
\frac{1}{\sqrt{2}}~{\rm diag}(1,0,-1)~,
~~~\frac{1}{\sqrt{2}}~{\rm diag}(0,1,-1)~,
\label{charges}
\ee
respectively. Here, the first two entries of the matrices correspond to
the 
unbroken SU(2). Since only the non-Abelian part is non-BPS, these
two monopoles are attracted by the Coulomb potential 
\be
-\frac{g^2}{16\pi l^2} ~~~ (l > \mpi) ~,
\ee 
where $l$ is the monopole separation. Here a factor of $-\frac{1}{4}$
is from the inner product of the 
non-Abelian part of (\ref{charges}), and the Abelian part is neglected
because it is approximately BPS under our mass conditions mentioned
before.\footnote{In this SU(3) example, the
assumption that the 
Abelian Higgs mass is small is important, because otherwise the
Abelian gauge force has a factor of $\frac{3}{4}$, which makes
the overall color singlet configuration unstable.} 
When the two monopoles stay inside the range
$\mpi$, we can approximate the near-BPS fields outside of the massive
cores by the superposition of two
SU(2) monopoles at positions {\boldmath $r$}$_1$ and {\boldmath
$r$}$_2$. This 
gives the
Higgs fields at {\boldmath $r$} as
\be
{\rm diag}( t_1, t_1 + \frac{1}{\sqrt{2}er_1}
+ \frac{1}{\sqrt{2}er_2}, t_3 - \frac{1}{\sqrt{2}er_1} - \frac{1}{ 
\sqrt{2}er_2} )
\ee
if there were no non-Abelian cloud and
\be
{\rm diag}(t_1 + \frac{1}{\sqrt{2}er_1}, t_1 +
\frac{1}{\sqrt{2}er_2}, t_3 - \frac{1}{\sqrt{2}er_1} -
\frac{1}{\sqrt{2}er_2})
\ee
with a minimal size non-Abelian cloud, where ${\rm diag}(t_1,t_1,t_3)$
is 
the vacuum and $r_i= | \br - \br_i |$
$(i=1,2)$. In the latter case, the non-Abelian 
field is cancelled at a length scale bigger than the monopole
separation $l$. Therefore,
under the potential (\ref{potential}), it is
energetically favored to have a minimal size non-Abelian cloud
surrounding the massive monopoles. However the non-Abelian
Higgs field is still present within the separation scale
$l$. Integrating 
(\ref{potential}) over the spatial region up to
$\mpi$, we obtain an 
attractive potential\footnote{For big $g$, this kind of forces will
affect the binding energies of the multiple monopole states considered
in \cite{Gardner:uu} where similar Higgs mass relations are taken.}
\be
\frac{g^2}{32\pi} m_{\phi}^2 l + {\cal O}(g^2 m_{\phi}^3 l^2) ~~~ 
(l < \mpi)~.
\label{lpotential}
\ee
So, if the core size is ignored, the massive monopoles will
coincide classically and have a 
non-Abelian cloud bound to them. This is similar to what we have seen
in SO(5).

\section{Non-Abelian monopole clouds and dual Meissner effect}
The energy scale $\mph$ and the linear property of the potentials
(\ref{apotential}) and (\ref{lpotential})
may receive corrections from the higher order terms neglected in
(\ref{potential}). However, the
following qualitative features do not depend 
on these terms and the specific examples. Within the ${\cal N}=4$ 
supersymmetry length scale $\mpi$ around the massive monopoles, the
appearance of the 
non-BPS Higgs raises the energy above the vacuum due to the non-BPS
potential; outside of this scale, we have the BNC
instability; so, whenever the topology is allowed, the non-Abelian
clouds 
will always contract to cancel the non-Abelian fields of the enclosed
massive monopoles.

In our discussion, because the massive monopoles
carry non-Abelian magnetic charges, they actually serve as probes
so that we can 
study the properties of the dual non-Abelian states. Unlike
the Coulomb-like phase in ${\cal N}=4$ as we saw in
Sec.~\ref{SecDuality}, these dual states
now have effective masses and the non-Abelian magnetic charges are
screened. In other words, in this intermediate energy region
where we describe the magnetic theory by solitons,
breaking the supersymmetry by a 
superpotential (but maintaining
the non-Abelian nature of the vacuum) in the weakly coupled electric
theory causes the magnetic theory to be 
in an analogous dual Higgs phase. In the following,
we will discuss the possibility of this phenomenon continuously
going to the dual Meissner effect when we lower the energy scale
$\mph$ to that of the vacuum state ($\Lambda_{QCD}$), where the test
massive solitonic monopole becomes the 
test elementary particle.

To do this, we first note that, although the ${\cal N}=1$ non-Abelian
vacuum has
the energy scale $\Lambda_{\rm QCD}$, we have only seen the
non-Abelian clouds at $m_{\phi}$ because we rely on the presence of
massive non-Abelian monopoles. To look at these non-Abelian clouds at
a lower energy scale $\tilde m_{\phi}$ ($m_{\phi} > \tilde m_{\phi} >
\Lambda_{\rm QCD}$) with a corresponding bigger electric coupling
$\tilde e$ (according to the asymptotic freedom), we should
change the set-up by lowering the supersymmetry 
breaking scale to $\tilde{m}_\phi$ and choose the ${\cal N}=4$ theory
above it to have the corresponding coupling $\tilde{e}$. By the same
argument we see that, after the supersymmetry breaking, the
non-Abelian clouds of the ${\cal N}=1$ theory with coupling
$\tilde{g}$ 
are Higgsed and get a mass $\sim \tilde{m}_\phi$. The same reasoning
can go all the way to $e\lesssim 1$ ($g \gtrsim 1$).    

From $e \sim g \sim 1$ around $\Lambda_{QCD}$, the solitonic
description we used in the magnetic 
theory starts to deviate from being a good approximation. For the
massive 
monopoles, the Compton wavelength begins to exceed the monopole core
size. For the non-Abelian clouds, the
potential becomes too shallow. 
Only one bound state can exist, with a mass gap $g^2 \mph$ determined
by the depth of the potential. This bound state has a wavelength of
order $g^{-4} \mpi$, which begins to exceed the range of the potential
$\mpi$ as $g \lesssim 1$. (Outside of $\mpi$, we still have the BNC
instability in the Higgs direction we are considering.) So as
mentioned before, below $g \sim 1$ we should 
switch the roles of 
elementary particles and the solitons between the electric and
magnetic theories.
These analyses also suggest that the masses (which should
be of order $\Lambda_{QCD}$ from the last paragraph) of 
the dual non-Abelian fields are likely to vary continuously,
rather than abruptly vanish, at $g \sim 1$.\footnote{For the solitonic
description of the magnetic theory at big
$g$, the massive monopoles can only probe one Higgs direction since
the non-Abelian clouds are non-zero in only one of the
Higgs fields. There it is 
enough that we have a superpotential for one chiral multiplet. But in
order for this screening effect to continuously go to the case $g
\lesssim 1$ where the non-Abelian Higgs can oscillate in all
directions, the 
superpotential in the other two complex directions of the Higgs should
also be present as we mentioned in footnote {\ref{ftHiggs}}.}

This meets the expectation that the usual weakly 
coupled dual Higgs mechanism starts to take effect. Nielson-Olesen
electric flux tubes \cite{Nielsen:cs} appear as solitonic objects and
this causes confinement of non-Abelian electric charge and electric
fields. The
quantum fluctuations of these tubes are of order $g$ times the
thickness of the flux tubes \cite{Nielsen:cs}. Here we comment that
for big $g$ above 
$\Lambda_{QCD}$, these fluctuations are much bigger than the size of
the electric flux. This is consistent with the fact
that the electric fields are not confined above $\Lambda_{QCD}$,
despite of the analogous dual Higgs mechanism.

The coupling stops running soon after  
the magnetic perturbation theory starts to become valid, since all
the non-Abelian magnetically charged particles obtain masses of order
$\Lambda_{QCD}$ through this dual Higgs mechanism.

Since so far all the non-BPS properties of the ${\cal
N}=1$ 
theory that we have used are shared by the non-supersymmetric theory,
we can further 
break the ${\cal N}=1$ supersymmetry by adding some 
soft breaking terms. For example, we can add a non-Abelian gaugino
mass term with mass equal to the supersymmetry breaking scale $\mph$ and 
get the same picture.

\acknowledgments 
I would like to thank Brian Greene, David Tong, Piljin Yi and
especially Dan Kabat and Erick Weinberg for many helpful discussions
and comments on the manuscript. This work was supported in part by the
U.S. Department of Energy.

\end{document}